\documentclass[floatfix,onecolumn,pre]{revtex4}
\usepackage[pdftex]{graphicx}
\usepackage{amssymb}
\usepackage{amsmath}
\usepackage{latexsym}
\usepackage{bm}
\usepackage{times,subfigure}
\newcommand{\gamSV}{\ensuremath{\gamma_{\mathrm{SV}}}}
\newcommand{\gamSL}{\ensuremath{\gamma_{\mathrm{SL}}}}
\newcommand{\young}{\ensuremath{\theta_{\mathrm{Y}}}}
\newcommand{\cassie}{\ensuremath{\theta_{\mathrm{C}}}}
\newcommand{\wenzel}{\ensuremath{\theta_{\mathrm{W}}}}

\begin{document}

\title{Surface evolver simulations of drops on microposts}

\author{MATTHEW L. BLOW{\footnote{Now at: Centro de F\'{i}sica Te\'{o}rica e Computacional (CFTC), University of Lisbon, Instituto de Investiga\c{c}\~{a}o Interdisciplinar, Av. Prof. Gama Pinto, 2, Lisboa, 1649-003, Portugal}} {\footnote{{\bf email:} matthewlblow@gmail.com}}}
\author{JULIA M. YEOMANS}

\address{The Rudolf Peierls Centre for Theoretical Physics, University of Oxford,\\
1 Keble Rd, Oxford, OX1 3NP, England}

\begin{abstract}

An important feature in the design of superhydrophobic surfaces is their robustness against collapse from the Cassie-Baxter configuration to the Wenzel state. Upon such a transition a surface loses its properties of low adhesion and friction. We  describe how to adapt the Surface Evolver algorithm to predict the parameters and mechanism of the collapse transition on posts of arbitrary shape. In particular, contributions to the free energy evaluated over the solid-liquid surface are reduced to line integrals to give good convergence. The algorithm is validated for straight, vertical and inclined, posts. Numerical results for curved posts with a horizontal section at their ends show that these are more efficient in stabilising the Cassie state than straight posts, and identify whether the interface first depins from the post sides or the post tips. 

\end{abstract}

\maketitle

\section{Introdution}

 A small liquid drop placed on the flat surface of a partially wetting material forms a spherical cap. A measure of the wettability of the surface is the contact angle $\young$, defined as the angle between a tangent to the drop and the substrate, and related to the surface tensions between the three coexisting phases by Young's equation~\cite{Youngref}
\begin{equation}
\cos\young=\frac{\gamSV-\gamSL}{\gamma}\;    \label{eqn:young}
\end{equation}
where $\gamSV$, $\gamSL$ and $\gamma$ are the solid-vapour, solid-liquid and liquid-vapour surface tensions.
Materials with contact angles $\young>90^\circ$ or $\young<90^\circ$ with respect to water are referred to as hydrophobic or hydrophilic respectively. 

If a surface is covered by micron-scale bumps or posts, which can be stiff or flexible, its wetting properties are altered. In general, when a hydrophobic surface is roughened the contact angle is increased, leading to superhydrophobic behaviour~\cite{Quere}. Conversely, the contact angle of a hydrophilic material decreases as the surface becomes rough on a micron length scale~\cite{BicoTordeux}. Micropatterned structures are relatively easy to fabricate using modern microlithography techniques, and superhydrophobic structures in particular are finding increasing applications as water repellent surfaces, energy efficient dehumidifiers or cooling devices\cite{dehumidifer}. A surprising number of plants and insects use superhydrophobic adaptations: examples include the water-repellent leaves of the lotus and nasturtium~\cite{Neinhuis} and hairs on the legs of the water strider~\cite{Gao}. 

\begin{figure}
\centering
\includegraphics[width=80mm]{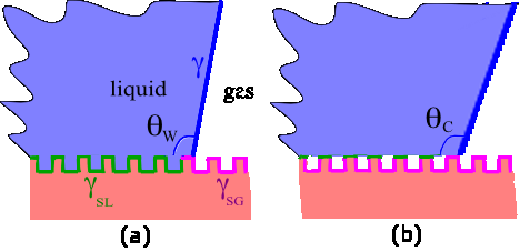}
\caption{(a) Collapsed and (b) suspended states on a superhydrophobic surface. $\wenzel$ and $\cassie$ are the Wenzel and Cassie-Baxter angles respectively, and the $\gamma's$ denote the surface tensions between adjacent phases.}
\label{fig:wettingStates}
\end{figure}

There are at least two possible states for a drop resting on a patterned surface. In the first of these, the collapsed or Wenzel state~\cite{Wenzel}, shown in Fig.~\ref{fig:wettingStates}(a), the fluid fills the interstices between the posts. The effective contact angle in the collapsed state is
\begin{equation}
\cos\wenzel=r\cos\young\; \label{eqn:wenzel}
\end{equation}
where the roughening of the substrate increases its surface area by a factor $r$. The Wenzel statel dominates the phase diagram for hydrophilic or weakly hydrophobic contact angles.

The second common state is the suspended or Cassie-Baxter state~\cite{Cassie}; the drop lies on top of the posts with pockets of air beneath it, as shown in Fig.~\ref{fig:wettingStates}(b). The effective contact angle in the suspended state is
\begin{equation}
\cos\cassie={\phi}\cos\young-(1-\phi)\; \label{eqn:cassie}
\end{equation}
where the projection of the liquid-solid contact area onto the base surface takes up an area fraction $\phi$. As expected the suspended state tends to be thermodynamically stable, or metastable, for larger $\theta_Y$ and more closely spaced posts, although if the posts overhang, it can be metastable even for hydrophilic contact angles~\cite{McHaleAqil,CaoHu,TutejaEtAl}.
In the suspended state, a very high contact angle may be produced even for modestly hydrophobic materials and there is low resistance to droplet motion, an important property in many applications. Hence ensuring the stability and robustness of the suspended state is an important consideration in the application of both fabricated and natural rough surfaces, and it is interesting to calculate how this is affected by system parameters.

For large drops, so that the interface under the drop is flat, and simple post geometries, the dependence of the collapse transition from the Cassie-Baxter to the Wenzel state on post dimensions and spacing and contact angle can be calculated analytically. However for more complicated surface geometries this quickly becomes impossible and numerical methods are required to understand how the instability to collapse occurs. Here we show how Surface Evolver, a public domain software package written by Brakke~\cite{Brakke,SurfaceEvolverManual} which is widely used to predict equilibrium interface configurations, can be extended to identify the collapse transition on surfaces patterned by arrays of posts of arbitrary shape. 

Surface Evolver calculates the shape of a surface which minimises a given functional, in our case the free energy. The surface is parametrised in terms of triangular facets which are sequentially refined to give increasingly good approximations to the optimum surface configuration. Restrictions on the position of the surface are provided by boundaries, here the positions of contact lines between the surface and the posts. Hence to calculate the interface shape and free energy for a drop on an array of posts we need to define a suitable parametrisation for the position of the posts and to write the free energy contributions in terms of that parametrisation.

We first describe how to parametrise post shapes, and then how to calculate and minimise the free energy of an interface in contact with the posts. Results are given for straight, inclined and bent posts; the first two of these can be compared with exact results to test the algorithm, the third can only be treated numerically. A discussion suggests possible applications of the approach.

\section{Model}

\begin{figure}
\centering
\includegraphics[width=120mm]{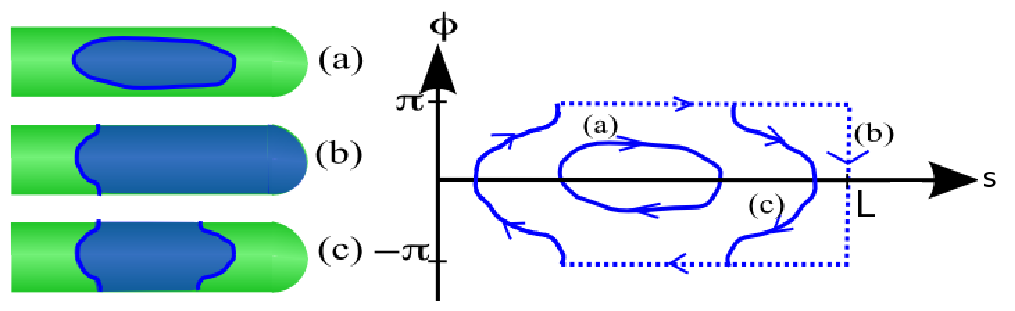}
\caption{Contact lines of various topologies (left), and the contours of their corresponding line integrals in $(s,\phi)$ (right), where the wetted area is enclosed by (a) a single contact line, not including the tip of the post, (b) a single contact line including the tip of the post, and (c) two separate contact lines wrapped around the post.}
\label{fig:boundary}
\end{figure}

We consider thin cylindrical posts, with bases located at $z=0$, arranged in a rectangular array with lattice vectors $D_{x}\mathbf{e}_{x}$ and $D_{y}\mathbf{e}_{y}$. The posts are modelled by constructing a central spine of length $L$, described by the curve $\mathbf{r}_{0}(s)$, where $s$ is an arclength parameter, such that $\vert\dot{\mathbf{r}}\vert=1$. The surface of the post is defined to lie at a specified perpendicular radius $a(s)$ away from the spine. Hence it can be parameterised by $s$ and an axial angle $\phi$ as
\begin{equation}
\mathbf{r}(s,\phi)=\mathbf{r}_{0}(s)+a(s)\left(-\mathbf{n}(s)\cos\phi+\mathbf{b}(s)\sin\phi\right)\;.    \label{eqn:3Dparameterisation}
\end{equation}
Here $\mathbf{n}$ is the unit normal to the spine in the direction of its curvature, such that
\begin{equation}
\ddot{\mathbf{r}_{0}}=\kappa\mathbf{n}\;,  \label{eqn:curvature}
\end{equation}
where $\kappa$ is the scalar curvature,
 and
\begin{equation}
\mathbf{b}=\dot{\mathbf{r}_{0}}\mathbf{\times}\mathbf{n}
\end{equation}
 is the unit binormal (${\dot{\mathbf{r}_{0}}},\mathbf{n},\mathbf{b}$ form the Frenet-Serret frame). We assume posts to have a constant radius, except for a rounded end-section, included to avoid sharp edges for computational convenience, such that
\begin{equation}\label{eqn:radiusFunction}
a(s)=\begin{cases}a_{0}\;, & 0<s<L-a_{0}\;,\\
\sqrt{a_{0}^{2}-(L-a_{0}-s)^{2}}\;, & L-a_{0}<s<L\;.\\ \end{cases} 
\end{equation}

We model the spine curve $\mathbf{r}_{0}(s)$ of the post using a piecewise function described by a discrete data set. One choice is to model the spine as a series of straight line segments. However, this does not work well in conjunction with an interface, as it leads to sharp corners between segments which can pin the interface. A better method is to discretise the post using segments which are circular arcs, which for simplicity we take to all have unit length. Under this discretisation scheme, the data needed to describe a post of length $N$ are the tangent vectors $\left\{\dot{\mathbf{r}}(p)\right\},\;p=0,1...N$ at each of the joins between segments.

The curvature of each segment is
\begin{equation}
\kappa(s)\equiv\kappa_{p,p+1}=\arccos\left[\dot{\mathbf{r}}(p).\dot{\mathbf{r}}(p+1)\right]\;,  \label{eqn:surfaceEvolverKappa}
\end{equation}
with $p=\lfloor s\rfloor$ (the largest integer smaller than $s$). The tangent vector within a segment is given by
\begin{equation}
\dot{\mathbf{r}}_{0}(s)=\frac{\dot{\mathbf{r}}(p)\sin\left[\kappa(p+1-s)\right]+\dot{\mathbf{r}}(p+1)\sin\left[\kappa(s-p)\right]}{\sin\kappa}\;,  \label{eqn:FrenetSerretTangent} 
\end{equation}
which may be integrated to find the position of the spine,
\begin{equation}
\mathbf{r}_{0}(s)=\mathbf{r}(p)+\frac{\dot{\mathbf{r}}(p)\cos\left[\kappa(p+1-s)\right]+\dot{\mathbf{r}}(p+1)\left(1-\cos\left[\kappa(s-p)\right]\right)}{\kappa\sin\kappa}\;.  \label{eqn:surfaceEvolverPosition}
\end{equation}
For a given segment, the binormal $\mathbf{b}$ is constant and is given by 
\begin{equation}
\mathbf{b}(s)\equiv\mathbf{b}_{p,p+1}=\frac{\dot{\mathbf{r}}(p)\mathbf{\times}\dot{\mathbf{r}}(p+1)}{\sin\kappa}\;,   \label{eqn:FrenetSerretBinormal}
\end{equation}
and the normal $\mathbf{n}$ is thus
\begin{equation}
\begin{split}
\mathbf{n}(s)&=\mathbf{b}(s)\mathbf{\times}\dot{\mathbf{r}}(s)    \\
&=\frac{-\dot{\mathbf{r}}(p)\cos\left[\kappa(p+1-s)\right]+\dot{\mathbf{r}}(p+1)\cos\left[\kappa(s-p)\right]}{\sin\kappa}\;.
\end{split}    \label{eqn:FrenetSerretNormal}
\end{equation}

We consider the array to be in contact with an interface spanning the $x$ and $y$ directions. For simplicity, we ignore edge effects and assume periodic boundaries so that the interfacial profile is the same for each post. The interface connects with the post at a contact line which, through Eqn.~(\ref{eqn:3Dparameterisation}), can be described by a relation between $s$ and $\phi$.


\section{Free Energy}
\label{sec:freeEnergy}

Next we write down the three contributions to the free energy of the system. The first is the interfacial energy between the water and air,
\begin{equation}
\mathcal{F}_{\mathrm{I}}
=\gamma\iint_{\mathcal{A}_{\mathrm{LG}}}dS .
\end{equation}
The liquid-gas interface ${\mathcal{A}_{\mathrm{LG}}}$ can easily be parametrised using facets within the Surface Evolver framework. An interfacial energy proportional to area is the default integral in the package, and the surface tension is chosen to be $1$. Using the value $\gamma_{\text{physical}}=0.072\mathrm{Nm^{-1}}$ for water, and taking one unit of length in our simulation to correspond to $1~\mathrm{\mu m}$, the pressure in the corresponding physical system is given through the Laplace relation as 
\begin{equation}
\Delta p_{\text{real}}=\frac{\gamma_{\text{physical}}}{\gamma_{\text{simulation}}}\frac{L_{\text{simulation}}}{L_{\text{physical}}}\Delta p_{\text{sim}}\approx 72\mathrm{kPa}\Delta p_{\text{sim}}.
\end{equation}
Since the micron scale is much smaller than the capillary length, any variation of Laplace pressure due to hydrostatics may be ignored. A suitable initial state is a horizontal interface, spanning the periodic boundaries in the $x$ and $y$ directions, and attached to the uppermost part of the hair.

The second contribution to the free energy is the wetting energy associated with the solid surfaces, here the surface of a post,
\begin{equation}
\begin{split}
\mathcal{F}_{\text{wetting}}&=\iint_{\mathcal{A}_{\mathrm{SL}}}\gamma_{\mathrm{SL}}(\mathbf{r})dS+\iint_{\mathcal{A}_{\mathrm{SG}}}\gamma_{\mathrm{SG}}(\mathbf{r})dS\\
&=\gamma\iint_{\mathcal{A}_{\mathrm{SL}}}-\cos\young(\mathbf{r})dS+\iint_{\mathcal{A}_{\mathrm{SL}}\cup\mathcal{A}_{\mathrm{SG}}}\gamma_{\mathrm{SG}}(\mathbf{r})dS\;,
\end{split}            \label{eqn:wettingEnergy}
\end{equation}
where $\mathcal{A}_{LS}$ and $\mathcal{A}_{LG}$ are the areas of the substrate in contact with liquid and gas respectively. Young's law~(\ref{eqn:young}) has been used in the second step to write the substrate tensions in terms of the equilibrium contact angle $\young$. Assuming that the total substrate area remains unchanged, the second term in the second line is a constant contribution to the free energy which may be ignored.

 Using Eqn.~(\ref{eqn:3Dparameterisation}), the expression for a cylindrical element of area on a post surface can be written
\begin{equation}
d\mathbf{S}(s,\phi)=a\left\{-\dot{a}\dot{\mathbf{r}}_{0}(s)+\left(1+a\kappa\cos\phi\right)\left(-\cos\phi\mathbf{n}(s)+\sin\phi\mathbf{b}(s)\right)\right\}ds d\phi\;.     \label{eqn:hairSurfaceElement}
\end{equation}
Thus, the free energy of the liquid wetting the post follows from Eqn.~(\ref{eqn:wettingEnergy}) as
\begin{equation}
\mathcal{F}_{\mathrm{W}}=-\gamma\cos\young\iint_{\mathcal{A}_{\mathrm{LS}}}a\sqrt{\dot{a}^2+\left(1+a\kappa\cos\phi\right)^{2}}dsd\phi\;.   \label{eqn:surfaceIntegralWetting1}
\end{equation}
$\dot{a}$ is zero except at the tip of the post. If we restrict $\kappa$ to be small at the tip, then the expression~(\ref{eqn:surfaceIntegralWetting1}) simplifies to 
\begin{equation}
\mathcal{F}_{\mathrm{W}}\approx-\gamma\cos\young\iint_{\mathcal{A}_{\mathrm{LS}}}a\sqrt{1+\dot{a}^{2}}\left(1+a\kappa\cos\phi\right)dsd\phi\;,
\end{equation}
and, using Eqn.~(\ref{eqn:radiusFunction}), gives
\begin{equation}
\mathcal{F}_{\mathrm{W}}\approx-\gamma\cos\young\iint_{\mathcal{A}_{\mathrm{LS}}}a_{0}\left(1+a\kappa\cos\phi\right)dsd\phi\;,
\label{eqn:surfaceIntegralWetting} 
\end{equation}
which conveniently holds for both regions $0<s<L-a_{0}$ and $L-a_{0}<s<L$.

${\mathcal{A}_{\mathrm{LS}}}$ is fixed rather than free and, as a result, it is not suitable to model this contact surface directly by facets, especially if the post is curved. This is because, for a hydrophobic substrate, there is a tendency for individual facets to grow or shrink in order to shortcut the substrate curvature, while for a hydrophilic substrate the facets may overlap one another or spread beyond the contact line. Thus the post is distorted away from its prescribed shape.

Therefore, the integration over $\phi$ is performed analytically, to give a line integral,
\begin{equation}
\mathcal{F}_{\mathrm{W}}=-\gamma\cos\young \oint_{\partial\mathcal{A}_{\mathrm{LS}}}a_{0}\left(\phi+a\kappa\sin\phi\right)ds\;. \label{eqn:boundaryIntegralWetting}
\end{equation}
This is evaluated numerically using the parameterised 'boundary' capability of Surface Evolver, with vertices specified in the parameter space of $(s,\phi)$ and then mapped to positions in Cartesian space by Eqn.~(\ref{eqn:3Dparameterisation}).

The boundary $\partial\mathcal{A}_{\mathrm{LS}}$ depends on the topology of the contact line, as shown in Fig.~\ref{fig:boundary}. In the case where the contact line loops around the trunk of the post (case (b) in the figure), there is a branch cut where $\phi$ jumps from $\pi$ to $-\pi$ and $\partial\mathcal{A}_{\mathrm{LS}}$ must continue to the end of the post at $s=L$ in order to include the wetted area beyond the end of the contact line. It is also possible that states with two contact lines, when the post is fully submerged in some parts, may arise. These may be accounted for by `gluing' adjacent edges together along the line $\phi=\pi$, in which case the contour integral is taken along the path shown in Fig.~\ref{fig:boundary}(c).

The third and final contribution to the free energy is the Laplace pressure,
\begin{equation}
\mathcal{F}_{\text{Laplace}}=-\Delta p\left(V_{\mathrm{L}}-V_{0}\right)\;,
\end{equation}
where $V_{\mathrm{L}}$ is the volume of the liquid phase, and $\Delta p$ is the difference in pressure between the liquid and gas phases, and $V_{0}$ is a constant. The Laplace pressure $\Delta p$ is either variable, in which case the volume is constrained as  $V_{\mathrm{L}}=V_{0}$ ($\mathcal{F}$ is then the Helmholtz free energy), or fixed, in which case $V_{\mathrm{L}}$ varies (Gibbs free energy). In either case $\Delta p$ induces a non-zero mean curvature of the surface.

The Laplace energy may be rewritten as a surface integral over the boundary of the fluid,
\begin{equation}
\mathcal{F}_{\text{Laplace}}=\Delta p\int\int_{\mathcal{A}_{\mathrm{LG}}\cup\mathcal{A}_{\mathrm{LS}}}z\mathbf{e}_{z}.d\mathbf{S}\;.    \label{eqn:laplaceIntegral}
\end{equation}
Within Surface Evolver, the portion of (\ref{eqn:laplaceIntegral}) taken over $\mathcal{A}_{\mathrm{LG}}$ is most easily evaluated as a vector integral over the surface, but for the same technical reasons as discussed for the wetting integral~(\ref{eqn:wettingEnergy}), the contribution from the solid surfaces is most efficiently evaluated as a line integral. Therefore we write
\begin{multline}
\mathcal{F}_{\mathrm{L}}=\Delta p\iint_{A_{\mathrm{LG}}}z\mathbf{e}_{z}.d\mathbf{S}   \nonumber   \\
+\iint_{\mathcal{A}_{\mathrm{LS}}}\left(z_{0}+a\left[-n_{z}\cos\phi+b_{z}\sin\phi\right]\right)\left(-\dot{a}a\dot{z_{0}}+a\left(1+a\kappa\cos\phi\right)\left[-\cos\phi n_{z}+\sin\phi b_{z}\right]\right)d\phi ds\;.
\end{multline}
and perform the integral over $s$ explicitly to give
\begin{multline}
\mathcal{F}_{\mathrm{L}}=\Delta p\iint_{\mathcal{A}_{\mathrm{LG}}}zdxdy +\Delta p\oint_{\partial\mathcal{A}_{\mathrm{LS}}}\Bigl(az_{0}\left[-n_{z}\sin\phi-b_{z}\cos\phi-\dot{a}\dot{z_{0}}\phi\right]\\
           +a^{2}[\tfrac{1}{2}\phi\left(n_{z}^{2}+b_{z}^{2}-\kappa n_{z}z_{0}\right)+\dot{a}\dot{z_{0}}\left(n_{z}\sin\phi+b_{z}\cos\phi\right)\\+\tfrac{1}{4}\left(2n_{z}-\kappa z_{0}\right)\cos2\phi+\tfrac{1}{4}\left(n_{z}^{2}-b_{z}^{2}-\kappa n_{z}z_{0}\right)\sin2\phi]\\
+a^{3}\kappa\left[\tfrac{1}{2}b_{z}n_{z}\left(\cos\phi+\tfrac{1}{3}\cos 3\phi\right)+\tfrac{1}{4}\left(3n_{z}^{2}+b_{z}^{2}\right)\sin\phi+\tfrac{1}{12}\left(n_{z}^{2}-b_{z}^{2}\right)\sin 3\phi\right]\Bigr)ds\;,
\label{eqn:boundaryIntegralPressure}
\end{multline}
which is again integrated over the boundaries shown in Fig.~\ref{fig:boundary}.


\section{Results}
\begin{figure}
\centering
\includegraphics[width=150mm]{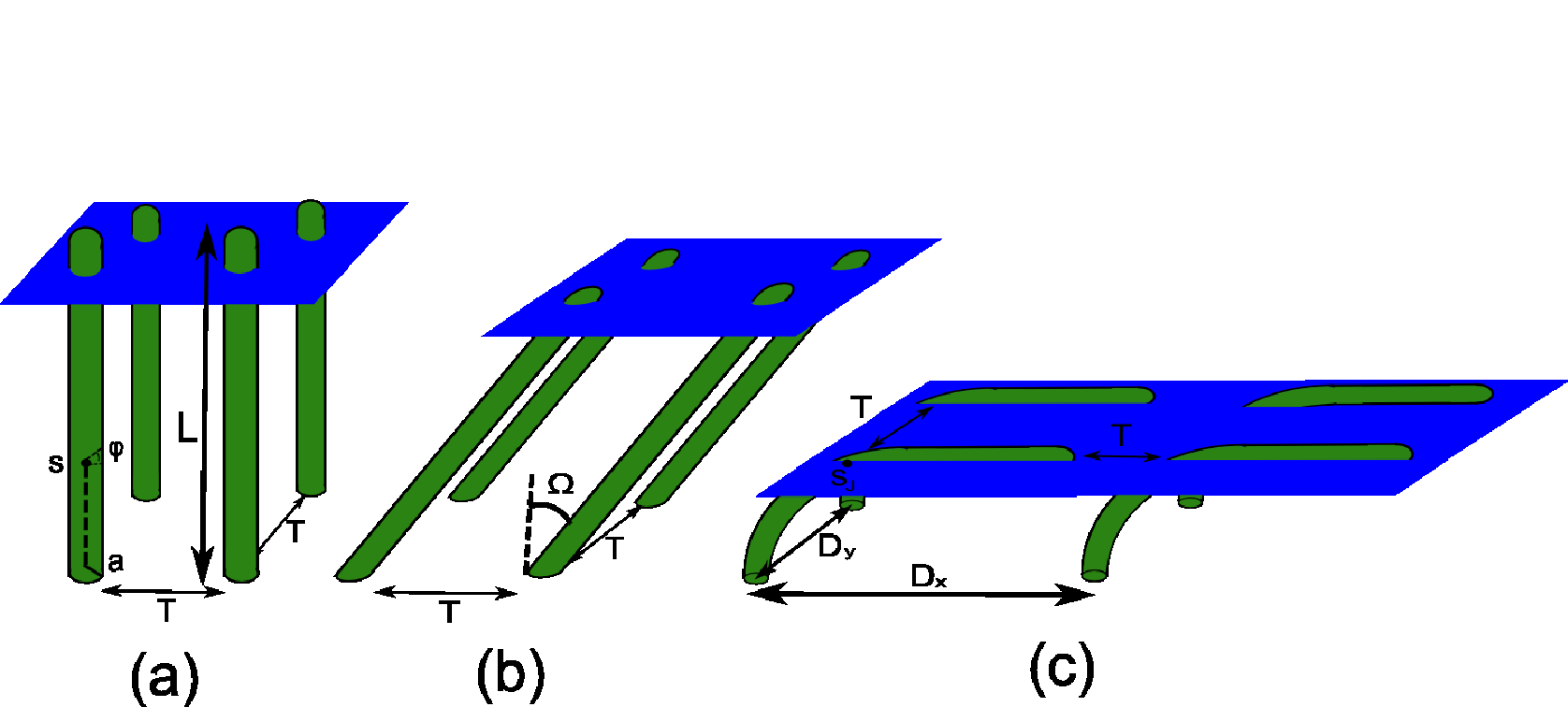}
\caption{Schematic illustration of the hair geometries considered: (a) Straight hairs perpendicular to the base substrate. (b) Straight hairs inclined at an angle $\Omega$. (c) curved hairs with a section parallel to the base substrate, described by Eqn.~(\ref{eqn:rShapedHairPos}). The diagrams indicate the position of the base of the drop in the Cassie state.}
\label{fig:hairDiagrams}
\end{figure}

We begin by considering the threshold for collapse from the Cassie-Baxter to the Wenzel state for straight vertical posts, as depicted in Fig.~\ref{fig:hairDiagrams}(a). The shape of the interface when a Laplace pressure is applied will be of a nontrivial form, but for this particular geometry, it need not be known to calculate the critical pressure for collapse $\Delta p_{\text{max}}$. This is because the system has a non-varying cross-section, and as such, an advance $-\delta z$ of the contact line will not change the shape of either the contact line or the interface. Therefore, the free energy change is~\cite{CrispThorpe}
\begin{equation}
 \delta \mathcal{F}=\left\{\gamma 2\pi a\cos\young + \Delta p\left(D_{x}D_{y}-\pi a^{2}\right)\right\}\delta z\;,      \label{eqn:freeEnergyPlastronStraight}
\end{equation}
and hence the interface will advance down the posts, leading to collapse, at a Laplace pressure
\begin{equation}
\Delta p_{\text{max}}=\frac{-2\pi a\gamma\cos\young}{D_{x}D_{y}-\pi a^{2}}\;.    \label{eqn:verticalCrit}
\end{equation}
Similarly, when the posts are straight but inclined at an angle $\Omega$ to the $z$ direction (Fig~\ref{fig:hairDiagrams}(b)), a generalisation of Eqn.~(\ref{eqn:verticalCrit}) is easily derived by noting that the effect of inclination is to shorten the perpendicular distance between posts~\cite{CrispThorpe}
\begin{equation}
\Delta p_{\text{max}}=\frac{-2\pi a\gamma\cos\young}{D_{x}D_{y}\cos\Omega-\pi a^{2}}\;. \label{eqn:inclinedCrit}
\end{equation}

\begin{figure}
\centering
\includegraphics[width=150mm]{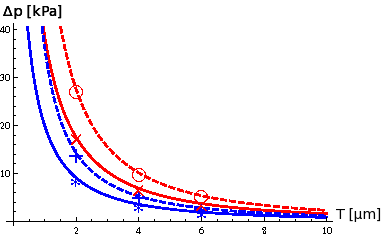}
\caption{Variation of depinning pressure with $D_{x}=D_{y}=T+2a$ for straight hairs, with theory (lines) and simulation results (markers) compared. Values are shown for the parmeters $\young=105^{\circ},\Omega=0^{\circ}$ (blue solid line, $*$ markers), $\young=105^{\circ},\Omega=45^{\circ}$ (blue dashed line, $+$ markers), $\young=120^{\circ},\Omega=0^{\circ}$ (red solid line, $\times$ markers), and $\young=120^{\circ},\Omega=45^{\circ}$ (red dashed line, $\circ$ markers),   }
\label{fig:straightHairsGraph}
\end{figure}

These results provide a useful validation of the algorithm. In Fig.~\ref{fig:straightHairsGraph} we plot the variation of the Laplace pressure for collapse with post spacing for a square lattice of posts, with $a=1$ and spacing $S$ between posts such that $D_{x}=D_{y}=T+2a$, for two representative contact angles. Each result is obtained by starting with the Laplace pressure at zero and increasing it in small increments; between each increase the surface configuration is iterated until equilibrium is reached. The final Laplace pressure at which the interface collapses, which corresponds to the limit of metastability of the suspended state, is then recorded.

 Fig.~\ref{fig:straightHairsGraph} shows that the Surface Evolver algorithm is able to reproduce the exact results well. As expected collapse is easier for less hydrophobic posts. Inclining the posts tends to stabilise the suspended state because the effective gap between the posts is reduced.

We also present results for posts which are curved. The lower section bends in a circular arc, until it is pointing horizontally at an arclength $s_{\mathrm{J}}$, at which point there is a straight horizontal section, as shown in Fig.~\ref{fig:hairDiagrams}(c). This profile resembles the hairs found on the plastrons of certain aquatic arthropods~\cite{ThorpeCrispA,FlynnBush,Bush}, and posts of this shape can be fabricated artificially~\cite{Kim,Chu,HsuSigmund}. The corresponding spine function is
\begin{equation}
\mathbf{r}=\begin{cases}\frac{2s_{\mathrm{J}}}{\pi}\left(1-\cos\tfrac{\pi s}{2s_{\mathrm{J}}}\right)\mathbf{e}_{x}+\frac{2s_{\mathrm{J}}}{\pi}\sin\tfrac{\pi s}{2s_{\mathrm{J}}}\mathbf{e}_{z}\;,   & 0\leq s\leq s_{\mathrm{J}}\;, \\
\left(\frac{2s_{\mathrm{J}}}{\pi}+s-s_{\mathrm{J}}\right)\mathbf{e}_{x}+\frac{2s_{\mathrm{J}}}{\pi}\mathbf{e}_{z}\;,                & s_{\mathrm{J}}\leq s\leq L\;.   \end{cases}     \label{eqn:rShapedHairPos}\;.
\end{equation}

\begin{figure}
\centering
\includegraphics[width=150mm]{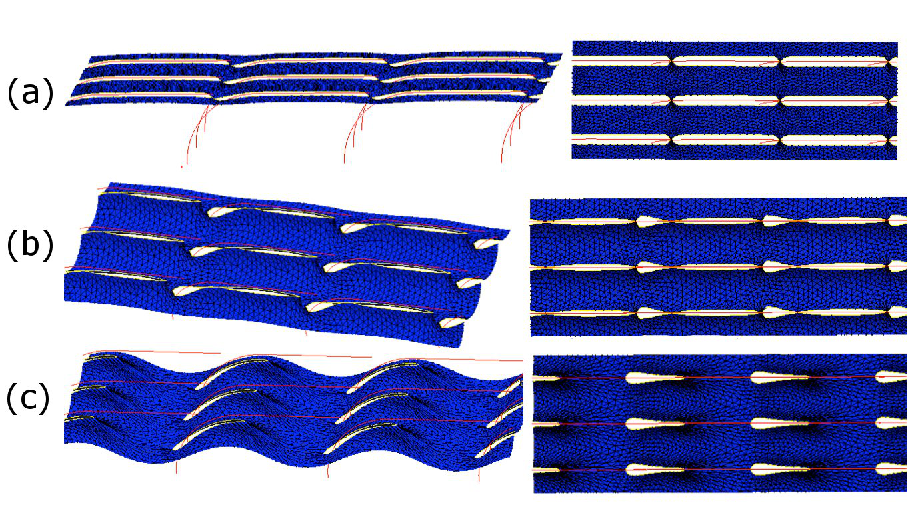}
\caption{Graphical output of the Surface Evolver simulation for curved hairs. The spines of the hairs are represented as red lines, and the interface is shown in blue: (a) a typical state where the interface is pinned to the horizontal sections of the hairs, (b) lateral depinning of the interface from the sides of the hairs, (c) tip depinning from the ends of the hairs.}
\label{fig:rShapedHairs}
\end{figure}

\begin{figure}
\centering
\includegraphics[width=150mm]{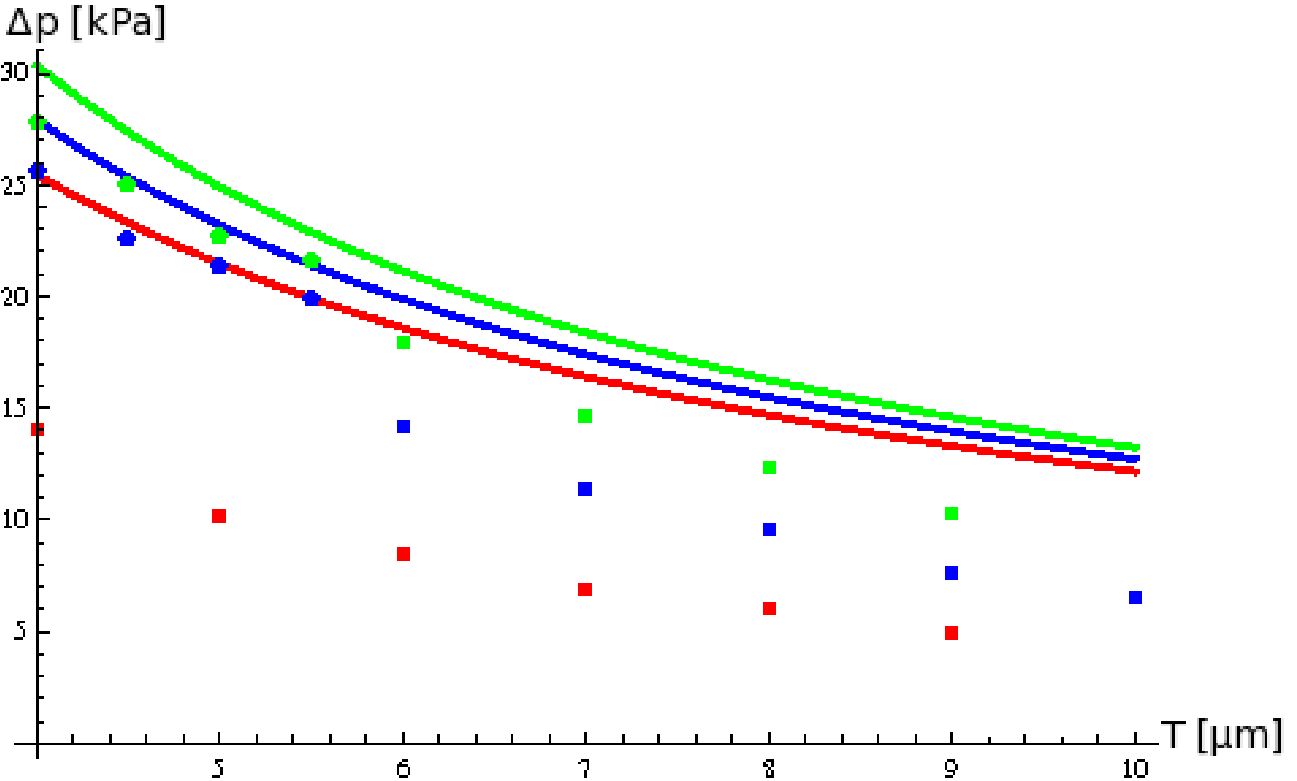}
\caption{Variation of depinning pressure with post spacing for curved hairs with $D_{x}=T+L_{x}-s_{\mathrm{J}}+2a$ and $D_{y}=T+2a$, for Young angles $\young=90^{\circ}$ (green), $\young=105^{\circ}$ (blue) and $\young=120^{\circ}$ (red). The shape of the symbols distinguishes lateral (circles) and tip (squares) depinning.}
\label{fig:rShapedHairsGraph}
\end{figure}

We choose $a=1$, $L=32$ and $s_{J}=16$. In the $y$ direction, we define the spacing between the posts, $T$, by $D_{y}=2a+T$. In the $x$ direction, there is a greater degree of ambiguity in the definition of the spacing. We choose $D_{x}=T+L_{x}-s_{\mathrm{J}}+2a$. Fig.~\ref{fig:rShapedHairsGraph} shows results for the threshold pressure as a function of spacing for three values of $\young$. The horizontal section of the posts tends to pin the interface (demonstrated in Fig.~\ref{fig:rShapedHairs}(a)), providing a means of stabilising the suspended state, even when the posts are not hydrophobic. 

The simulations also allow us to identify the path by which the interface moves down the posts, thus identifying the most vulnerable feature of a given post design. For smaller $T$ and more hydrophobic $\young$, collapse occurs when the interface depins laterally, around the barrel of the hair, as shown in Fig.~\ref{fig:rShapedHairs}(b). The threshold pressure for lateral depinning was calculated by Chrisp and Thorpe \cite{CrispThorpe} to be
\begin{equation}
\frac{\Delta p_{\text{max}}}{\gamma}=\begin{cases}
\left[\tfrac{1}{2}\sqrt{D_{y}^{2}-\left(2a\sin\young\right)^{2}}-2a\cos\young\right]^{-1}\;,&\cot\young<\frac{a}{2D_{y}}\;.    \\
\tfrac{2\sin\young}{D_{y}}\;,&\cot\young>\frac{a}{2D_{y}}\;,\end{cases}  \label{eqn:lateralCrit}
\end{equation}
This result is compared to the simulations in Fig.~\ref{fig:rShapedHairsGraph}.  Eqn.~(\ref{eqn:lateralCrit}) assumes a system which is uniform in the $x$ direction (essentially a row of infinitely long hairs), which leads to the prediction being slightly higher than the simulation results. Fig.~\ref{fig:rShapedHairs}(b) shows that, for finite hairs, lateral depinning occurs around the 'crook' of the hair at $s=s_{\mathrm{J}}$, not uniformly along the section $s_{\mathrm{J}}<s<L$ as assumed in the derivation of Eqn.~(\ref{eqn:lateralCrit}).

When the hairs are not hydrophobic or when the spacing is sufficiently large, the interface depins from the tip, rather than the sides,  of the hair, as depicted in Fig.~\ref{fig:rShapedHairs}(c). This occurs at significantly lower threshold pressure than Eqn.~(\ref{eqn:lateralCrit}). For the cases $\young=105^{\circ}$ and $\young=120^{\circ}$, a sharp drop is observed in the graph Fig.~\ref{fig:rShapedHairsGraph} as $T$ is increased, marking the switching from lateral to tip depinning.

\section{Discussion}

An important feature in the design of superhydrophobic surfaces is their robustness against collapse from the suspended, Cassie-Baxter configuration to the collapsed, Wenzel state. Upon such a transition a surface loses its properties of low adhesion and friction. Here we have described how to extend the Surface Evolver algorithm to predict the parameters and mechanism of the collapse transition on posts of arbitrary shape. In particular, contributions to the free energy evaluated over the solid-liquid surface need to be reduced to line integrals to give good convergence. 

We have tested the algorithm against the exact results available for straight, vertical and inclined, posts. Numerical results for curved posts with a horizontal section at their ends show that these are more efficient in stabilising the Cassie state, and identify whether the interface first depins from the post sides or the post tips. 

Several insects have superhydrophobic adaptations to allow them to live at length scales where surface tension dominates. For example the legs of the water strider are covered with hairs to help it to float on the water surface and to leave the water easily as it jumps. The water spider has a body covered with stiff hairs to trap a bubble of air, the plasteron, which it uses for respiration\cite{Stratton}.  Moreover, devices have been constructed which exploit meniscus-dominated buoyancy. These include microboats made from wire mesh\cite{PanWang,Swinburne}, and a robot that mimics a water strider\cite{robot}. The approach presented here will help to investigate the relation of the detailed shape of the hairs to biological function and hence in the design of such biomimetic devices. 
Indeed numerical modelling will be important in designing optimal micropatterned surfaces for a wide range of applications such as drag reduction, water harvesting and water repellency.

\section*{Acknowledgments}
We thank Z. Guo and T. Swinburne for useful discussions.

\end{document}